# Spatial alignment, group strategy and non-kin selection enable the evolution of cooperation

Xiaoliang Wang[1,2]*, Andrew Harrison[3]*

1 College of Life Sciences, Zhejiang University, Hangzhou 310058, China
2 School of Physical Sciences, University of Science and Technology of China, Hefei 230026, China
3 Department of Mathematical Sciences, University of Essex, Colchester CO4 3SQ, UK

*Correspondence: (X.W.) wxliang@mail.ustc.edu.cn; (A.H.) harry@essex.ac.uk;

**Abstract**

This article considers a mechanism to explain the emergence and evolution of social cooperation. Selfish individuals tend to benefit themselves, which makes it hard for the maintenance of cooperation between unrelated individuals. We propose and validate that a smart group strategy can effectively facilitate the evolution of cooperation, provided cooperators spatially align whilst cooperating at a new level of alliance. The general evolutionary model presented here shows that a non-kin selection effect is a possible cause for cooperation between unrelated individuals and highlights that non-kin selection may be a hallmark of biological evolution.

## Main text

A long-standing puzzle in biological evolution is how cooperative behavior can be maintained within the selfish natural world [1-6] (in Prisoner's dilemma, defection is often the evolutionarily stable strategy, ESS, see Fig. 1). Explaining cooperation is essential to understand the major evolutionary transitions in biology [7]. Many explanations for cooperation [8-27] are proposed, such as kin selection [4,8,9], group selection [10-13], 'tit-for-tat' strategy [14-16], reciprocity [17-19], policing and punishment [4,20], and biological range expansion [21-23]. However, the basic evolutionary question: why the living world evolves from simple to complex organizations, has not been explained directly.

Kin selection theory, which is encapsulated in Hamilton's rule: $rb - c > 0$ ($r$ is the genetic relatedness between actor and recipient, $b$ and $c$ are benefits and costs of cooperation, respectively), argues that cooperation will evolve if the benefits produced by an individual's cooperative behavior lead to individuals with genes that increase their inclusive fitness [9,28]. Such an explanation can only account for the maintenance of cooperation among genetic relatives, especially in animal societies [29-31]. Although many theories and models have also been proposed for cooperation between non-kin [32-34], there are few unified evolutionary models and explanations for the continued cooperation between kin and that between non-kin, which has universally occurred in cooperation between genomes, organelles and even humans [28,32,35]. A unified evolutionary theory is important as it can help to reveal the evolutionary rules for all organisms. Here, based on a quantitative agent-based approach [36,37], we propose that a smart group strategy (GS), taken by a group of individuals, independent on whether they are genetically related or unrelated, can effectively increase their fitness and promote their competition with opponents. We develop a general evolutionary model to explain the preservation of cooperation between related and unrelated individuals. We find that non-kin selection is a natural law of biological evolution.

The idea of a group strategy arises from our counterintuitive observation: chemotaxis of microbial individuals, the directed motion guided by beneficial chemical cues with which individuals can improve their fitness, impedes, rather than promotes, their competition with opponents at the population level (Figs. 2a and 2b). This highlights a conflict between individual and population-level interests (Figs. 2c and 2d). This result demonstrates that in the competition between populations, an individual's evolutionary advantage might not be that of a population. Instead, it may even become deleterious to a population's evolution, and the short-sighted selfish behaviors of individuals may not be that favored as we expect. This situation sets a higher requirement for the evolution of populations, and a smarter group strategy will be favored.

We first validate the group strategy (see Box 1 for methods and the game rule) for the evolution of cooperation, through implementing it into the natural selection term of the stepping stone model (SSM, see Supporting Information S1.1). Fig. 1a shows that when cooperators take group strategy in the Prisoner's dilemma taking place within a one-dimensional closed space, cooperation is increasingly favored and is finally selected for evolution, with the higher levels of GSs taken (larger $\mathcal{L}$, which means larger cooperation alliances involving more members). This result is important for evolutionary theory, which has showed the possibility of natural selection in explaining the self-organization of individuals into complex biological organizations in the cheating world. In theory (the well-mixed finite population), the condition for the evolutionary stability of cooperation (cooperation is ESS) is derived as:

$$\sum_{k=1}^{\mathcal{L}} f_{g,k} e_{g,k} > b + c - d + \frac{d-b}{f_0} \quad (1)$$

where $e_{g,k}$ is the benefit obtained via the group strategy which is conducted among $k$ cooperators, $f_{g,k}$ is the frequency that the total individuals conducting that group strategy occupy within the cooperator population, $\mathcal{L}$ represents the highest GS level taken, and $f_0$ is the initial frequency of cooperators within the whole population.

We consider the linear benefit distribution of GSs (Fig. 1b). With the higher benefit increment produced by a new level of GS (higher α), cooperators can take a lower level of GSs to outcompete defectors (Fig. 1c), which suggests the importance of cooperation efficiency. The difference between simulations and the theory (Box 1 Eq. 2) is mainly due to the finite population size effect [38]. From the obtained phase diagram in the two-dimensional space consisting of $\mathcal{L}$ and the population size $N$ (population density) (Fig. 1d), we can observe the evolution phase of cooperation invades the non-evolution phase at small $N$, but stabilizes at a steady interface when $N > 40$.

We further include the effect of the initial frequency $f_0$, and plot in Fig. 1e the landscape of the lowest GS level $\mathcal{L}_w$ taken for the evolution of cooperation. We can see that both higher $N$ and $f_0$ can facilitate the evolution of cooperation, with the lower required $\mathcal{L}_w$ to be taken. In particular, the evolution of cooperation is impossible even for a high initial frequency, provided the population size is too small (e.g. $N < 10$). This means that a sufficient number of individuals are necessary to establish an effective group strategy.

> ### Box 1: Group strategy
>
> Consider species A (cooperators) within a population takes the group strategy (GS), namely establishes a cooperative spatial alignment within a group of independent individuals with any genotypes. Let $\mathbb{G} = [g_1, g_2, \cdots, g_k, \cdots, g_{\mathcal{L}}]$ denote a set of group strategies conducted within finite populations, where $g_k$ is the group strategy conducted among $k$ individuals (Box figure a), and $\mathcal{L}$ represents the highest cooperation level of the system ($\mathcal{L} \leq N$). Note that $k = 1$ indicates no cooperation between individuals.
>
> Let $N_{g,k}$ be the number of $g_k$, then the number distribution of group strategies $\mathbb{N}_{\mathbb{G}} = [N_{g,1}, N_{g,2}, \cdots, N_{g,k}, \cdots, N_{g,\mathcal{L}}]$. Let $n_{g,k} = k \cdot N_{g,k}$ be the total number of individuals that participate into $g_k$, then the size distribution of the populations involved in specific group strategies $n_{\mathbb{G}} =$ $[n_{g,1}, n_{g,2}, \cdots, n_{g,k}, \cdots, n_{g,\mathcal{L}}]$, with $\sum_{i=1}^{\mathcal{L}} n_{g,k} = n_A$, where $n_A$ is the total number of species A inside the whole population. Let $f_{g,k} = n_{g,k}/n_A$ be the frequency that the total individuals conducting $g_k$ occupy within the whole population of species A, then the frequency distribution $f_{\mathbb{G}} = [f_{g,1}, f_{g,2}, \cdots, f_{g,k}, \cdots, f_{g,\mathcal{L}}]$, with $\sum_{i=1}^{\mathcal{L}} f_{g,k} = 1$. Let $e_{\mathbb{G}} = [e_{g,1}, e_{g,2}, \cdots, e_{g,k}, \cdots, e_{g,\mathcal{L}}]$ denote the benefit distribution of the set of group strategies $\mathbb{G}$, where $e_{g,k}$ is the benefit that an individual can obtain via the group strategy $g_k$, so we can derive the resultant fitness of one cooperator within this system as:

$$a = f_{\mathbb{G}} \cdot e_{\mathbb{G}} = \sum_{k=1}^{\mathcal{L}} f_{g,k} e_{g,k}$$

where $a$ in the payoff matrix has incorporated the cooperation structure (different levels of cooperation among individuals). When $\mathcal{L} = 1$, $a = e_{g,1}$, which represents the inherent fitness of cooperators resulting from their individual cooperative behavior.

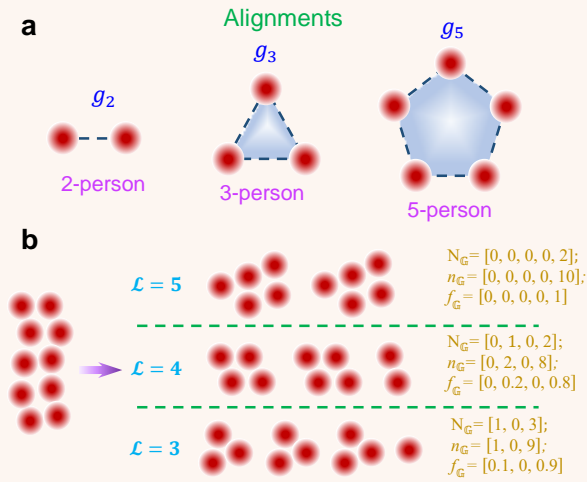

**Box figure.** (**a**) Illustration of cooperative spatial alignments taking the specific group strategy by cooperators. (**b**) The game rule in SSM simulations.

According to the Nash equilibrium (the well-mixed infinite population), the condition for cooperation is ESS is:

$$w_{A,0} = af_0 + b(1 - f_0) > w_{B,0} = cf_0 + d(1 - f_0)$$

Namely,

$$\sum_{k=1}^{\mathcal{L}} f_{g,k} e_{g,k} > b + c - d + \frac{d-b}{f_0} > c \quad (1)$$

In our SSM simulations, we make cooperators preferentially take the high-level group strategy, provided there is a sufficient number of individuals to compose that group (the game rule). For example, for a 10-person group, if their highest cooperation level is to take the group strategy $g_5$ (i.e. $\mathcal{L} = 5$), then these 10 individuals will organize into two 5-person alignments (Box figure b ). If $\mathcal{L} = 3$, the set of GSs taken by the 10 people are three 3-person alignments and one 1-person group.

For infinite populations, under our game rules, cooperators can all take the highest level of group strategy $g_{\mathcal{L}}$ if they succeed in evolving. In such a case, $f_{g,\mathcal{L}} \approx 1$. Therefore, in the well-mixed infinite population with the incorporation of the group strategy, the approximate condition for the evolution of cooperation is reduced to:

$$e_{g,\mathcal{L}} > b + c - d + \frac{d-b}{f_0} \quad (2)$$

In evolution, the group strategy can also be taken by defectors. One can expect that such a situation will enhance the difficulty for the evolution of cooperation, and compel cooperators to evolve to a higher level of cooperation organization. Within this expectation, cooperators always need to take a new level of group strategy when defectors take an increasingly high level of GS (Fig. 1f). This result can explain why the living world generally evolves in the direction of more complex organizations and the group strategy may play a significant role in biological evolution. Under more severe Prisoner's dilemmas (larger c-a and d-b), higher levels of GSs are expected to be taken (Fig. 1g).

We validated the effectiveness of group strategy for repeated games between cooperator and defector groups in well-mixed finite populations (Figs. 3a and 3b, see Box 2 for methods). We wondered how cooperation can be maintained between genetically unrelated individuals. To get the essence of problem, we develop a general model for the evolution of social cooperation (Fig. 4a). We find that cooperation can be preserved to a large degree among unrelated individuals, only if cooperators can be continuously supplemented with new cooperator members of genotypes that have little genetic dependence on old cooperator members (i.e. little kin selection effect).

When new cooperator members are selected from the gene pool with a strong genetic dependence on old members (the extreme example is that the new members are reproduced by old members), the total number of cooperator genotypes decreases over time, despite the continuous increase of cooperators within the whole population (Fig. 4b). After a long period of evolution, only one of the original fifteen cooperator genotypes survived and took over the whole population (genetic relatedness = 1) (Fig. 4c), with the others all removed during this process (Fig. 4d). However, if new cooperator members are supplemented with no genetic dependence on the cooperator group (e.g. new members are selected from the gene pool with the equal probability of the 15 genotypes), the cooperator group can still preserve those original 15 genotypes and maintain the low genetic relatedness among members over a long period of evolution (Figs. 4e-g).

We next investigate the effect of kin selection strength $C_{kin}$ on the evolution of cooperator genotypes in the population. With the reducing kin selection impact, more cooperator genotypes can be preserved in the cooperator group (Fig. 4h and Fig. 3c). We can further observe, from the landscape of the number of preserved cooperator genotypes $n_{gt}$ in the population, that the kin selection effect is stronger at smaller $N$ (below the dashed line), with fewer genotypes to be preserved (Fig. 4i). This is owing to the strong sampling effect arising from the small number of individuals at the low population size [37,39,40]. The impact of kin selection on $n_{gt}$ in a large range of $N$ is observed to obey a unified threshold function of the form (Fig. 5a):

$$n_{gt} = \frac{a(1 - C_{kin})}{(1 - C_{kin}) + (1 - K_{kin})} + 1 \quad (2)$$

where both the parameter $a$ and the threshold $K_{kin}$ are related to $N$ (Figs. 5b-d). Especially, $1 - K_{kin}$ is observed to exponentially decay as $N$ with a characteristic constant $\tau \approx 37$.

Our results demonstrate that the cause for the low genetic relatedness between individuals in the biological systems we have studied is the non-kin selection effect on various cooperator groups which is taken by the nature. Since natural non-kin selection is the evolutionary force (Box 2 Eq. 1), it will gradually drag the biological attribute of organisms from kin selection to the final non-kin selection. The most significant difference between humans and animals may be that we humans have evolved to abandon the absolute kin selection to survive throughout the operation of our lives (Fig. 4j).

> **Box 2: Evolutionary model for repeated social games**
>
> In each round of the game between cooperator and defector groups (Fig. 4a), the winning side has a chance to add one new member from the gene pool into its group, and the losing side will remove one old member from its group. This updating continues until one side completely takes over the whole population. The chance to add one new member for both sides is proportional to the current fitness of these two competition groups. The fitness of each side is calculated according to the payoff matrix described in the stepping stone model (see Supporting Information S1.1): $w_A = af + b(1-f)$, $w_B = cf + d(1-f)$. Here, $f$ is the frequency of cooperators in the whole population. Both sides can choose to take the group strategy proposed here. In that case, the element $a$ will be replaced with $a = f_\mathbb{G} \cdot e_\mathbb{G}$.
>
> To explore how cooperation can long be preserved among genetically unrelated individuals, we make the population continue to update even after cooperators have taken over the whole population. During this process, in each round, one new cooperator member is selected from the gene pool, and one old cooperator member in the group is removed.
>
> Throughout the whole evolutionary process, the removed member is randomly selected from its group with the probability proportional to its abundance in that group, while the newly added member is chosen randomly from the gene pool with the probability expressed as:
>
> <span style="color:green">**Biological kin selection**</span>     <span style="color:green">**Natural non-kin selection**</span>
>
> $$P_{gety,i} = C_{kin} f_{gety,i} + (1 - C_{kin}) \frac{1}{N_{gety}} \quad (1)$$
>
> $$\sum_{i=1}^{N_{gety}} P_{gety,i} = 1$$
>
> where $f_{gety,i}$ is the frequency of genotype $i$ in the cooperator/defector group, $N_{gety}$ is the total number of genotypes in the corresponding gene pool. $C_{kin} \in [0, 1]$ is the coefficient which can be used to tune kin selection strength. When $C_{kin} = 0$, the newly added member is selected with the equal probability of genotypes in gene pool that is genetically independent of old members in competition groups. When $C_{kin} = 1$, the addition of new members will completely depend on corresponding old genetically related members.

The group strategy proposed here is possibly a key evolutionary mechanism for shaping complex biological organizations. When and only when the non-kin selection effect is included, the cooperation between unrelated individuals can long be preserved, providing a good explanation for cooperation between non-kin.

**Acknowledgements:** This work is supported by Zhejiang University, National Natural Science Foundation of China and the research builds on research on the calibration of gene expression experiments funded by the UK BBSRC (BB/E001742/1). We are grateful to comments from Antonio Marco on an earlier version of the manuscript.

**Author contributions**: X.W. conceived this research, developed models, implemented theoretical calculations and data analysis, and wrote the paper. A.H. evaluated the article, provided revision suggestions and contributed to the writing. All authors participated in discussion.

**Conflict of interest**: The authors declare no conflict of interest.

**Data and materials availability**: All calculation data and code are available from X.W. upon request.

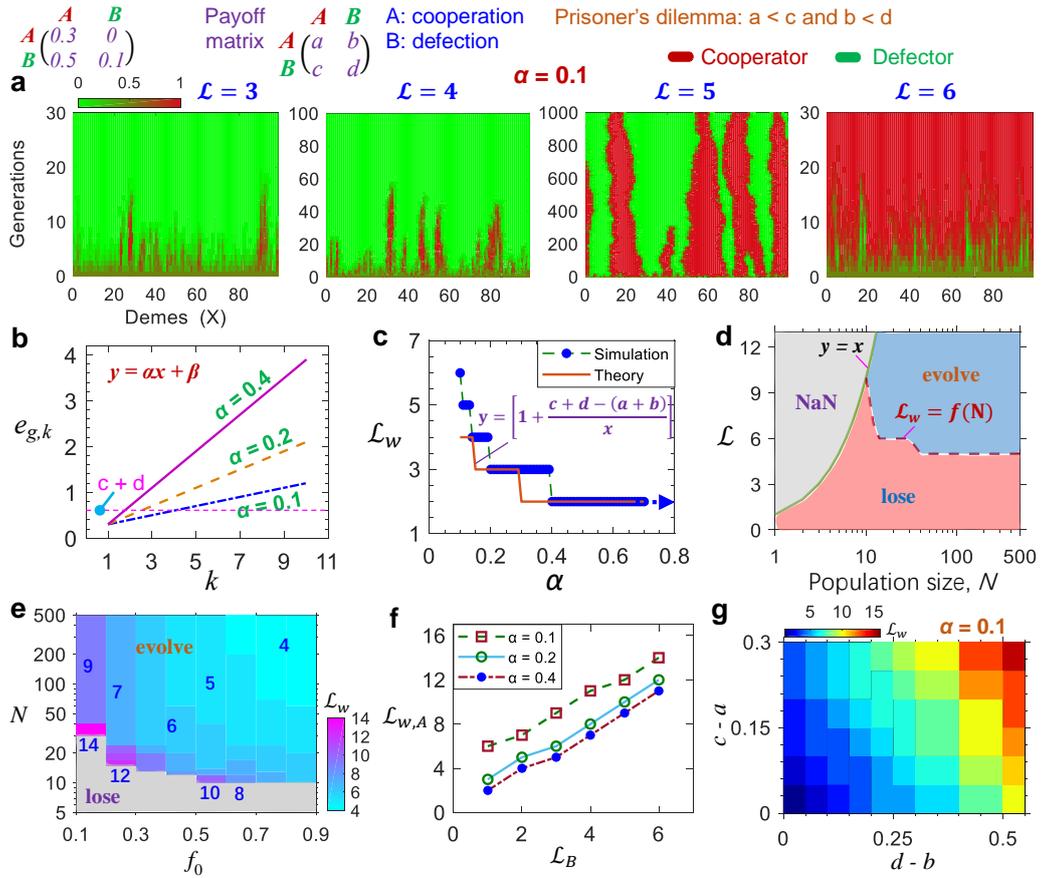

**Figure 1. Group strategy (GS) for the evolution of cooperation in Prisoner's dilemma within one-dimensional finite populations.** (**a**) Evolution of cooperation under different GS levels $\mathcal{L}$ (initial frequency $f_0 = 0.5$, $\alpha = 0.1$, population size $N = 30$). (**b**) Variation of benefit of the group strategy with the GS level. (**c**) The lowest GS levels taken for the evolution of cooperation under various benefit-elevating gradients $\alpha$ ($f_0 = 0.5$, $N = 30$). (**d**) The phase diagram for the evolution of cooperation ($\alpha = 0.1$, $f_0 = 0.5$). (**e**) Heat map of the lowest GS levels taken for the evolution of cooperation for varying population size $N$ and initial frequency $f_0$ ($\alpha = 0.1$). (**f**) The lowest GS level taken by cooperators for survival under different levels of GSs taken by defectors ($f_0 = 0.5$, $N = 30$). (**g**) Heat map of the lowest GS levels for varying c-a and d-b ($f_0 = 0.5$, $N = 30$, $\alpha = 0.1$). Elements (benefits) $a$ and $d$ are respectively the inherent growth rates of species A and B; $b$ is the additional fitness B imposed on A and $c$ is that A imposed on B. Both $b$ and $c$ only arise from social interactions between A and B. Default benefits in the payoff matrix are respectively 0.3 h$^{-1}$, 0, 0.5 h$^{-1}$ and 0.1 h$^{-1}$. Migration number mN = 2 per generation. Periodic boundary conditions are used. Each data point is tested for 10 rounds.

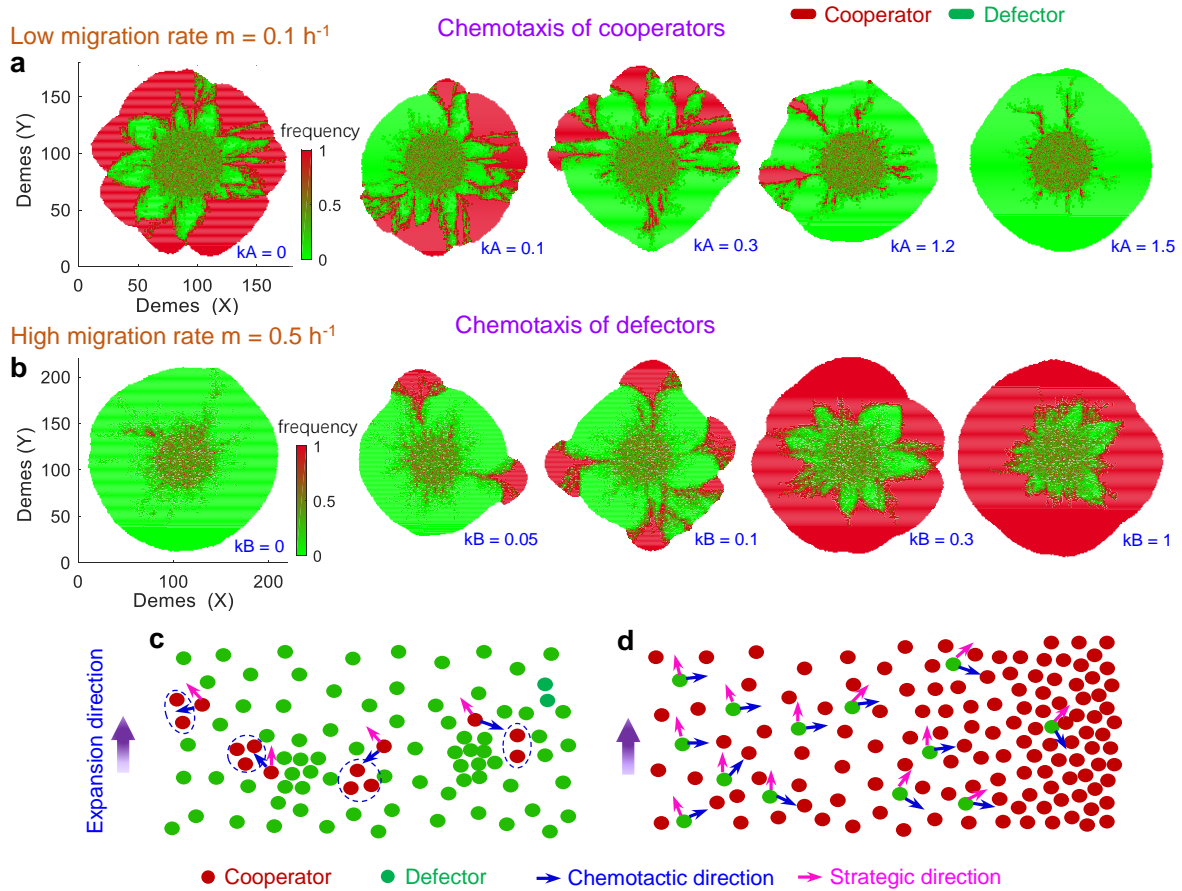

**Figure 2. 2D on-lattice simulation of the radial microbial range expansion (*E. coli* strains): chemotaxis effect on the evolution of cooperation in Prisoner's dilemma.** (**a**) Evolutionary games with the incorporation of chemotaxis of single cooperators and lower migration rate of individuals m = 0.1 h$^{-1}$: Cooperation can be selected in biological range expansion if the spatial segregation between cooperators and defectors can be established in a timely manner [19] (e.g. under the low migration rates, which means less frequent encounters with defectors). When the chemotaxis of cooperators is incorporated, with an increase in chemotaxis strength, evolutionary outcome is observed to transition from cooperation to defection. (**b**) Evolutionary games with the incorporation of chemotaxis of single defectors and higher migration rate of individuals m = 0.5 h$^{-1}$: Defection is selected under a higher migration rate of individuals, when no chemotaxis is involved. With the incorporation of chemotaxis of defectors and the increase in chemotaxis strength, natural selection increasingly favors cooperation. (**c**), (**d**) Illustration of the discrepancy between individual and population-level profits at the expanding frontier: (**c**) Chemotaxis of cooperators: Cooperators might move into a deme (location) where no defectors exist, but several cooperators might exist, which reduces the probability of them colonizing virgin space in advance. In considering the Prisoner's dilemma, however, the evolutionary advantage for cooperation is to colonize virgin space in the range expansion direction. Cooperation may not be favored by competition in such cases. (**d**) Chemotaxis of defectors: Defectors are more likely to move into the demes with more cooperators, which will leave aside a minor of cooperators which are at low density to survive and reproduce. Defectors will be always directed by cooperators in this case and are therefore in passive position in evolution. Parameters $k_A$ and $k_B$ represent the chemotaxis strength, the population size of each deme *N* = 20. Growth rates in the payoff matrix are respectively 0.3 h$^{-1}$, 0, 0.5 h$^{-1}$ and 0.1 h$^{-1}$. See Supporting Information for methods and the precision of our models.

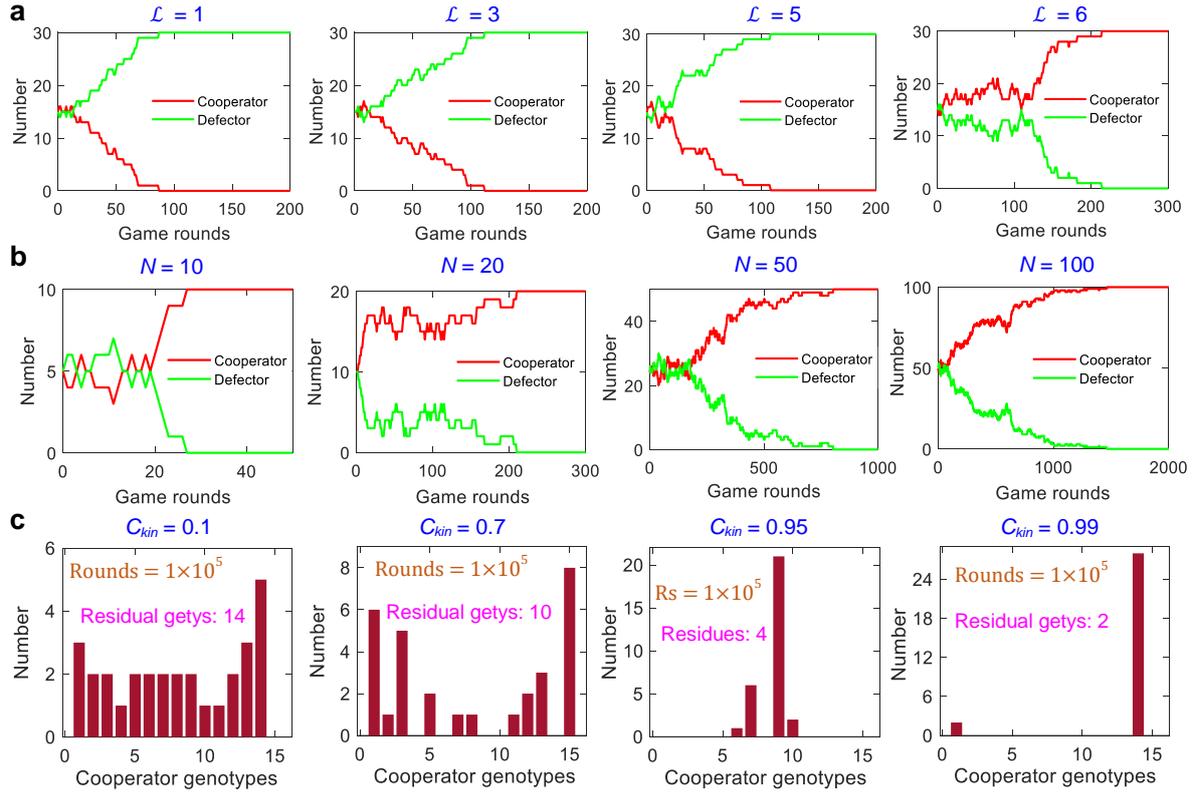

**Figure 3. Group strategy and non-kin selection effect in well-mixed finite populations.** (**a**) Cooperation is increasingly favored, with the higher level of group strategy $\mathcal{L}$ taken. Initial frequency $f_0$ = 0.5, benefit gradient $\alpha$ = 0.1, and the population size $N$ = 30. (**b**) Effect of the population size on the evolution of cooperation: Cooperation can be maintained within a large range of population size $N$. Initial frequency $f_0$ = 0.5, $\alpha$ = 0.1, $\mathcal{L}$ = 6. (**c**) Effect of kin selection strength $C_{kin}$ on the terminal number distribution of cooperator genotypes in the whole population (initial number of cooperator genotypes in the population is equal to that in the gene pool $N_{gety}$ = 15): With the increase in kin selection impact, fewer cooperator genotypes are preserved after a long period of evolution. $f_0$ = 0.5, $\alpha$ = 0.1, $\mathcal{L}$ = 6, $N$ = 30.

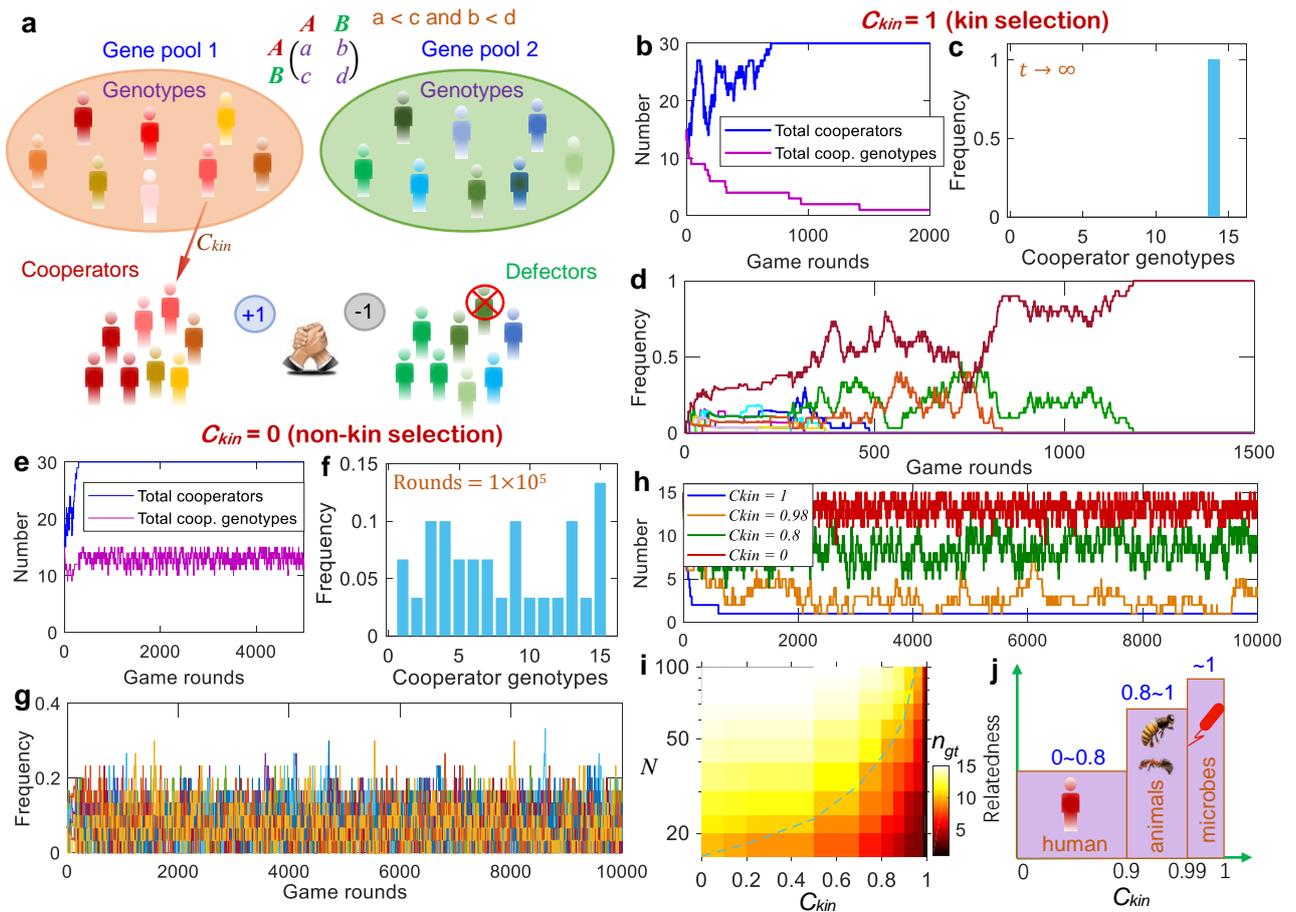

**Figure 4. The evolution of social cooperation in well-mixed finite populations.** (**a**) Illustration of the evolutionary model for repeated social games. (**b**)-(**d**) Kin selection effect: (**b**) Evolution of total numbers of cooperators and their genotypes with the number of game rounds. (**c**) The final frequency distribution of cooperator genotypes. (**d**) Evolution of the frequency distribution of cooperator genotypes within the cooperator group. (**e**)-(**g**) Non-kin selection effect: (**e**) Evolution of total numbers of cooperators and their genotypes. (**f**) The frequency distribution of cooperator genotypes at the game rounds = $1\times10^5$. (**g**) Evolution of the frequency distribution of cooperator genotypes. (**h**) Effect of kin selection strength $C_{kin}$ on the evolution of the total number of cooperator genotypes. (**i**) Heat map of the number of preserved cooperator genotypes $n_{gt}$ for varying population size $N$ and kin selection strength $C_{kin}$. Each data point is tested for 5 rounds. (**j**) Qualitative categories of kin selection effect in the living world. The default population size $N = 30$, initial cooperator frequency $f_0 = 0.5$, and the total number of cooperator genotypes in the gene pool $N_{gety} = 15$. Elements (scores) in the payoff matrix are respectively 0.3, 0, 0.5 and 0.1. Only cooperators take the group strategy (GS), and the GS level taken here $\mathcal{L} = 6$ ($\alpha = 0.1$).

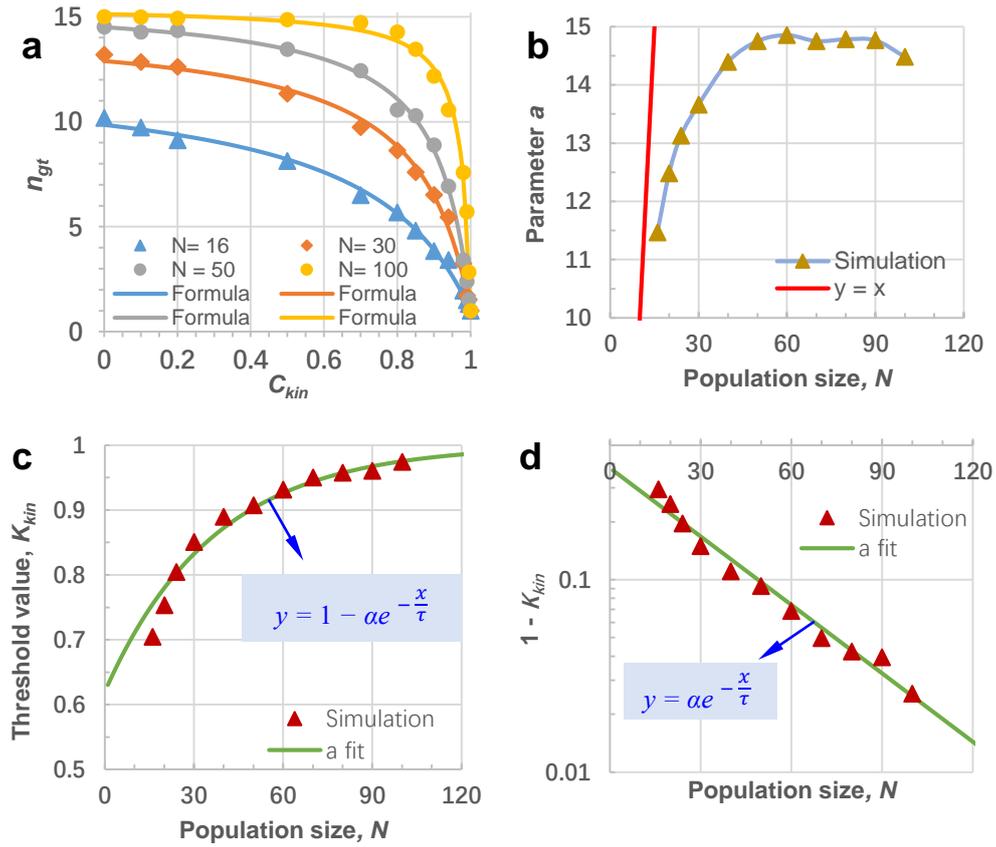

**Figure 5.** (**a**) The impact of non-kin selection on the number of preserved cooperator genotypes is observed to obey a simple threshold function law within a large range of population size $N$: Once kin selection impact $C_{kin}$ is beyond a certain threshold $K_{kin}$, there will be a prominent reduction in preserved cooperator genotypes (each data point is tested for 5 rounds). (**b**) Variation of the parameter $a$ in the threshold function with the population size $N$. (**c**), (**d**) Variation of the threshold value $K_{kin}$ that corresponds to prominent kin selection impact as a function of $N$: Plotted on (**c**) linear and (**d**) semi-logarithmic axes. $1 - K_{kin}$ is observed to exponentially decay as $N$. $\tau$ is the characteristic population size. $\alpha = 0.3798$, $\tau = 36.67$.